\documentclass[11pt, preprint]{aastex}
\usepackage{natbib}
\newcommand{\el}{$\ell$}
\newcommand{\gone}{PY Vul}
\newcommand{\Hb}{H$_\beta$}
\newcommand{\Hg}{H$_\gamma$}

\begin{document}
\title{The Peculiar Pulsations of PY Vul~}

\author{Susan E. Thompson,\altaffilmark{1} J. C. Clemens,\altaffilmark{1,2} 
M. H. van Kerkwijk,\altaffilmark{3} M. Sean O'Brien,\altaffilmark{4}\\
and D. Koester\altaffilmark{5}}

\altaffiltext{1}{Department of Physics and Astronomy, University of North Carolina 
Chapel Hill, NC 27599-3255; sthomp@physics.unc.edu, clemens@physics.unc.edu}
\altaffiltext{2}{Alfred P. Sloan Research Fellow}
\altaffiltext{3}{Department of Astronomy and Astrophysics, University of Toronto,
60 St. George Street, Toronto, ON, M5S 3H8, Canada}
\altaffiltext{4}{Astronomy Department, Yale University, PO Box 208101, 
New Haven, CT 06520-8101, USA}
\altaffiltext{5}{Institut f\"{u}r Theoretische Physik und Astrophysik, Universi\"{a}t
Kiel, 24098 Kiel, Germany}

\begin{abstract}
The pulsating white dwarf star \gone\ (G~185-32) exhibits pulsation
modes with peculiar properties that set it
apart from other variable stars in the ZZ Ceti (DAV) class.
These peculiarities include a low total pulsation amplitude, a 
mode with bizarre amplitudes in the ultraviolet, and a mode harmonic
that exceeds the amplitude of its fundamental. Here, we present optical, 
time series spectroscopy of \gone\ acquired with the Keck II 
LRIS spectrograph. Our analysis has revealed that the mode with
unusual UV amplitudes also has distinguishing characteristics
in the optical. Comparison of its line
profile variations to models suggests that this mode has
a spherical degree of four.  We show that all the other
peculiarities in this star are accounted for by a dominant pulsation
mode of \el=4, and propose this hypothesis as a solution to the mysteries 
of \gone.

\end{abstract}

\section{Introduction}

While each member of the variable DA white dwarfs (DAVs) 
shows its own unique set of pulsations, this group of pulsators
share many similar characteristics.
They all have a pure hydrogen atmosphere and 
reside in a narrow temperature strip near 
11,500~K, making them the coolest known class of white dwarf pulsators.
The multi-periodic brightness variations of these stars
are due to non-radial, g-mode pulsations \citep{RKN}
with periods between 
100~s and 1000~s and amplitudes less than a few percent. 
Those pulsators near the blue end of the instability
strip tend to show fewer modes with shorter periods and smaller
amplitudes than the pulsators near the red edge \citep{WF82}. 

  Understanding the characteristics of the pulsations
of individual DAVs provides the opportunity to model 
their interiors.
The ability to model the pulsations is limited
by our ability to identify the modes.
Pulsations are described by
the spherical degree (\el), azimuthal order ($m$), 
and radial order ($n$).
Unfortunately, the paucity of observed modes in DAVs
creates an obstacle to using period spacings for
mode identification.  Yet, some successful
mode identification has been performed on DAVs.
Long data sets from a single 
site, or using the Whole Earth Telescope (WET) 
has resolved azimuthal splittings, 
yielding the identification of \el\ and $m$ for
a few DAVs.  Time-resolved spectroscopy
in both the UV and optical wavelengths, has been 
applied to the brighter DAVs \citep[see][]{R95,VK00,Ke00,K02b,K02,T03,K03}.
This method identifies the spherical degree by measuring how
the amplitude of the mode changes with wavelength. 
Here, we apply the technique of optical, time-resolved
spectroscopy to the bright DAV, \gone\ and attempt
to decipher its prominent pulsation modes.

\paragraph{About \gone}

One of the brightest known DAVs (V$=12.97$ mag) with 
some of the smallest amplitude ($<0.3\%$) pulsations
is \gone\ (G~185-32, WD~1935+276).
Since its discovery as a pulsator \citep{M81}, it
was noted as having a curious pulsation spectrum. 
Atypical of small amplitude pulsators, \gone\ displays 
a wide range of periods, including
prominent modes near 370~s, 300~s, 215~s, 142~s, 72~s and 71~s
\citep{M81, Ke00, WET}. 
Given the star's shorter periods, low amplitudes,
relatively stable modes, and temperature,
this star is grouped with the pulsators near the blue edge
of the instability strip. 

The pulsation amplitudes on \gone\ are much lower than expected 
for DAVs of similar periods. DAVs follow a distinct trend:
those with larger amplitude modes have larger mean periods \citep{C94}.
The trend created by these stars (see Figure~\ref{davs}) reflects their
similarities and follows the expectations of mode driving 
mechanisms \citep{wg99,B92,Wi82}.  
\gone\ remains the exception to this trend.
Both its average mode amplitude and the amplitude of its largest mode
are approximately a factor of ten smaller than
the other DAVs. Suggested explanations for the low 
amplitudes have included nonlinear pulsation modes with a 
relatively large number of surface nodes \citep{M81}, a 
large magnetic field limiting the growth of its modes 
\citep{C94}, and a large inclination causing cancellation 
of the modes \citep{TC03}.

The mode at 71~s, the harmonic of the 142~s mode, 
is a curious feature of \gone. First, the presence of large
harmonic modes is normally reserved for the large amplitude pulsators.
Second, the amplitude of this harmonic has an amplitude similar
to the 142~s mode and has been observed to occasionally exceed the 
parent mode \citep[see][]{M81}; all other DAVs show harmonics
consistently smaller than the parent mode. The non-linear effects in the 
outer layers of DAVs, believed to be responsible for 
the presence of harmonics, should be small 
for low amplitude pulsations \citep{B92, Wu01, IK}. 
Some possible explanations for why a low amplitude
pulsator might display harmonics
were discussed in general by \citet{IK}
in their study of nonlinear effects on pulsations.
Nonlinear effects could appear with small amplitude pulsations 
if the star has an unusual surface convection zone,
 the star has a large inclination, or the mode has a large 
spherical degree (\el).

The most recent addition to the mysteries of 
the pulsations of \gone\ came from time-resolved UV
spectroscopy from the Hubble Space Telescope (HST) 
\citep{Ke00}. They observed that each 
of the modes' relative amplitudes increased in the 
UV as an \el\ $= 1$ or $2$ mode except for the 142~s mode.  
It shows no 
appreciable increase in amplitude while its harmonic at 71~s
still resembles a mode of \el $\le 2$. \citet{Ke00} suggests
that the 142~s mode is a result of nonlinear
mixing while the other modes, including the 71~s mode, 
are real pulsations.

\paragraph{The scope of this paper}
In this paper, we add our analysis
of optical time-resolved spectroscopy which offers an
appealing explanation for all the mysterious observations 
of \gone's modes. Our analysis of variations in the \Hb\ 
and \Hg\ lines show that the 142~s mode behaves like \el=4.
We therefore propose that this mode is
the dominant mode in the star and that it has a spherical
degree of four.
This hypothesis can also account for 
the low amplitudes, the UV characteristics 
of the 142~s mode, and the presence of its large 
harmonic mode.
We begin in \S\ref{data} by presenting
time series spectra and discussing
the variations of both the flux and velocity
variations measured from the spectra.
We then measure the pulsation amplitudes at each wavelength
across the spectra, present a new method
to perform this analysis, and compare the results
with the models.
In \S3 we propose that this star is dominated by a mode 
with \el=4 and show how this hypothesis is consistent
with the observations. We
discuss other explanations for the modes of
\gone\ in \S4 and finally, we summarize our conclusions
in \S5.

\section{Time-Series Spectroscopy}
\label{data}
\begin{figure}[!h]
\plotone{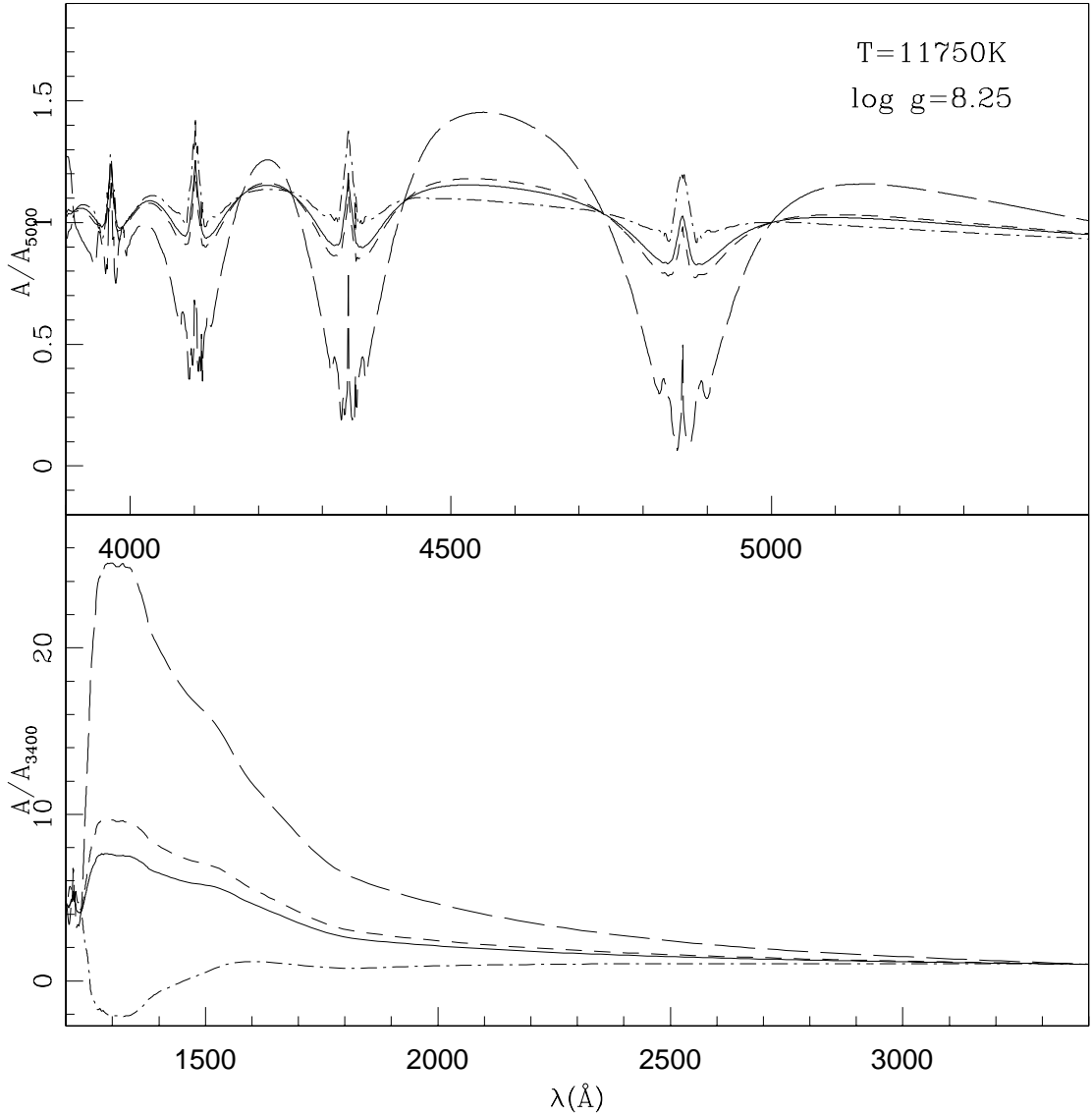}
\caption{Models of relative amplitudes for \el=1,2,3,4 
(solid, short dash, long dash and dot-dash) created using models
for a star at T=11,750 and log(g)=8.25.  Top panel covers the optical wavelengths
and has been normalized at 5000~\AA.  The bottom panel covers the UV and
are normalized at 3400~\AA.} 
\label{model}
\end{figure}

By analyzing the light curves at different wavelengths,
we can learn about the distribution of a pulsation 
mode across the surface of a star. \citet{R95} introduced this 
type of analysis for DAVs by using broad-band observations 
in the UV to measure the wavelength dependent amplitudes, 
thereby identifying \el\ of individual pulsation modes
in G~117-B15A.
This method uses the wavelength dependence 
of limb darkening and different cancellation effects to 
distinguish between different spherical degrees.  Figure~\ref{model}
shows model relative amplitude variations across the optical
and ultraviolet wavelengths for the first four spherical
degrees.

Van Kerkwijk, Clemens, \& Wu (2000) extended this technique
to optical, time-resolved spectroscopy of
the bright large amplitude DAV, G~29-38 and 
measured periodic shifts in the location of the spectral lines
associated with the motion of the stellar material during the
g-mode pulsations. \citet{C00}
measured the wavelength dependence of the pulsation amplitudes
of each mode (hereafter ``chromatic amplitudes'') and showed that
with the presence of velocities an asymmetry occurs
in the shape of the chromatic amplitudes.  
Since then, optical time-resolved spectroscopy
has been applied to HS~0507+0434B,
HL~Tau 76, and G~117-B15A \citep{K02, K02b, K03}.  
By comparing these results to model chromatic amplitudes similar to
the one presented in Figure~\ref{model}, these studies were able
to distingish some \el=2 modes from \el=1 modes.  

   The brightness of \gone\ suggested to us that similar observations
of this star would be productive.  In the following section, we discuss 
observations acquired with the Keck II LRIS spectrograph.
The analysis of our low-resolution, time-resolved spectra shows
that the signal-to-noise of the spectra is too low to see 
line profile variations in the chromatic amplitudes. 
However, we will introduce a new analysis
technique that enables us to extract information 
about the spherical degree from the series of spectra.

\subsection{Observations}

We observed \gone\ for two nights with the Low Resolution Imaging 
Spectrometer \citep[LRIS;][]{Oke} at the Keck II 
Observatory. On August 12, 1997 we took 410 spectral
images between 06:33:15.35 and 10:12:08.17 U.T.; on August 13, 1997
we took 396 images between 06:00:52.95
and 09:32:41.40 U.T.; times are as recorded by the image headers.
The times in these
headers are known to have substantial errors but
the timing intervals are accurate enough for our analysis.
The images collected during this 7.4 hours of data 
had an exposure time of 18~s with a read-out
time of approximately 14~s, achieved by reading the images 
through two amplifiers, binning the chip by two in the spatial 
direction, and reading only 50 rows of the CCD.  We
used the 600 line mm$^{-1}$ grating and the 8.7'' wide long slit such
that the seeing on the first night of 0.9'' resulted 
in a resolution of 4.9~\AA. On the second night
our seeing was 0.7'' yielding a resolution of 3.9~\AA. 
On each night, the series of exposures included images
of the Hg-Mg-Ar lamp for wavelength calibration and
of the flux standard Feige 110.

We reduced the spectral images by removing the bias, 
measured from the over-scan regions of 
both amplifiers and adjusting for the difference in
gain of the two amplifiers. We did not apply
flat fields because we did not obtain enough 
images to sufficiently reduce the stochastic noise 
in the average flat; application of the flat 
only increased the noise of the spectra.  
Using the apall routine of IRAF \citep[Imaging
Reduction and Analysis Facility,][]{To86} 
we traced and extracted the spectra from these images
removing the sky background and rejecting cosmic rays.
We applied the wavelength calibration and the 
flux calibration images to each spectrum.   
The extracted spectra cover a wavelength range 
from 3800 to 5700~\AA\ with a dispersion of  1.23~\AA~pixel$^{-1}$.
Measured at the continuum near 5000~\AA, the signal-to-noise of an individual 
spectrum is $\sim180$ per pixel while the average spectrum has a 
signal-to-noise of $\sim2200$ (See Figure 2). The
signal-to-noise of the average is less than expected from poisson statistics;
possibly because of sensitivity variation between pixels 
normally reduced by flat fielding.

\begin{figure}
\plotone{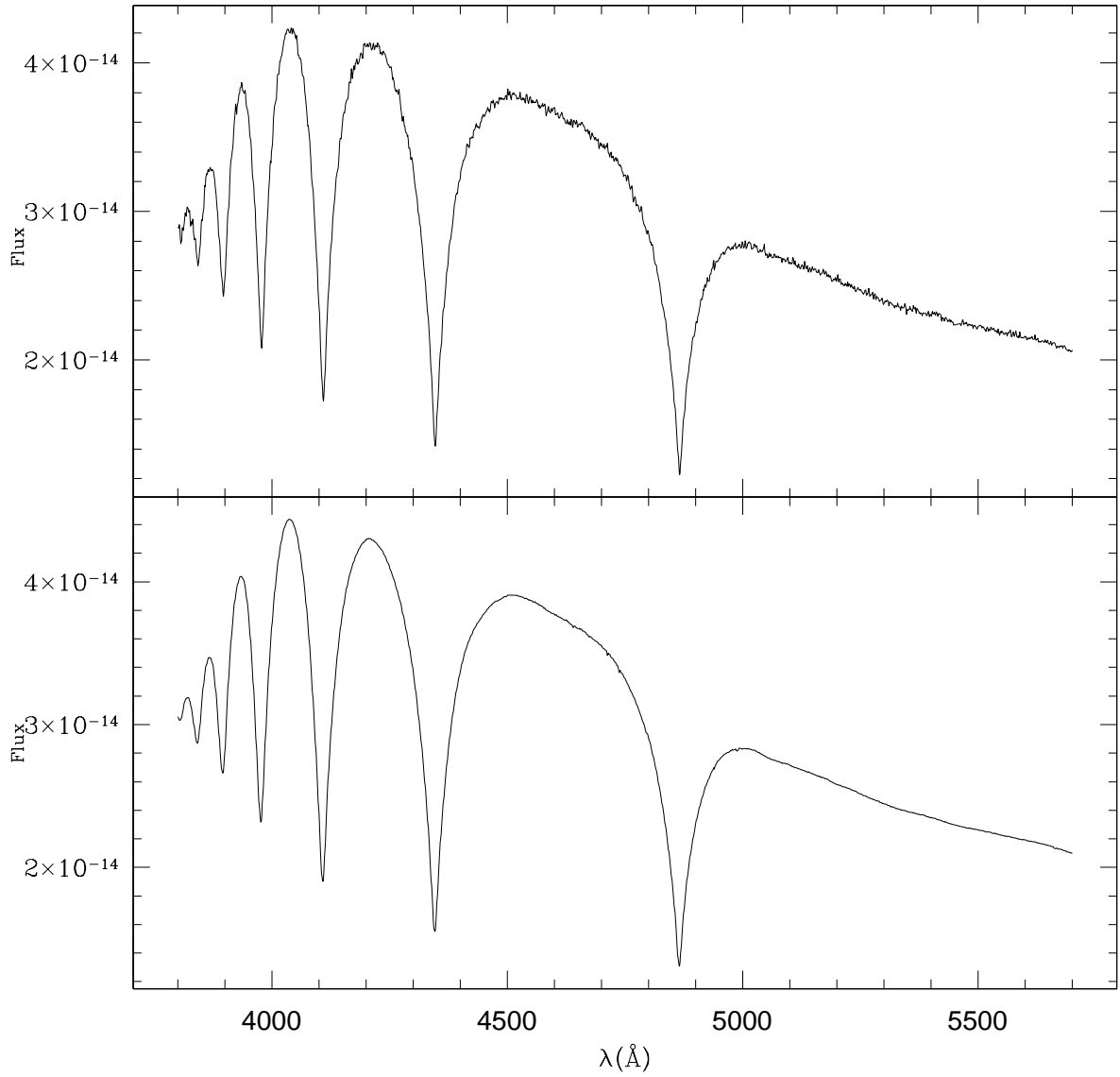}
\caption{An individual (top panel) and the average (bottom panel) reduced 
spectra of \gone\ taken on August 13, 1997.} 
\label{spec}
\end{figure}

\subsection{Light and Velocity Curves}
     Time-resolved spectroscopy provides the opportunity to
measure two aspects of a non-radial mode's pulsation, the brightness 
and the line profile variations.  The variations in the shape 
of the line are seen as both wavelength shifts and changes 
in the depth of the line \citep[see][]{C00}.
In the following section, we set out to extract
this information from the spectra of \gone.

\begin{figure}
\plotone{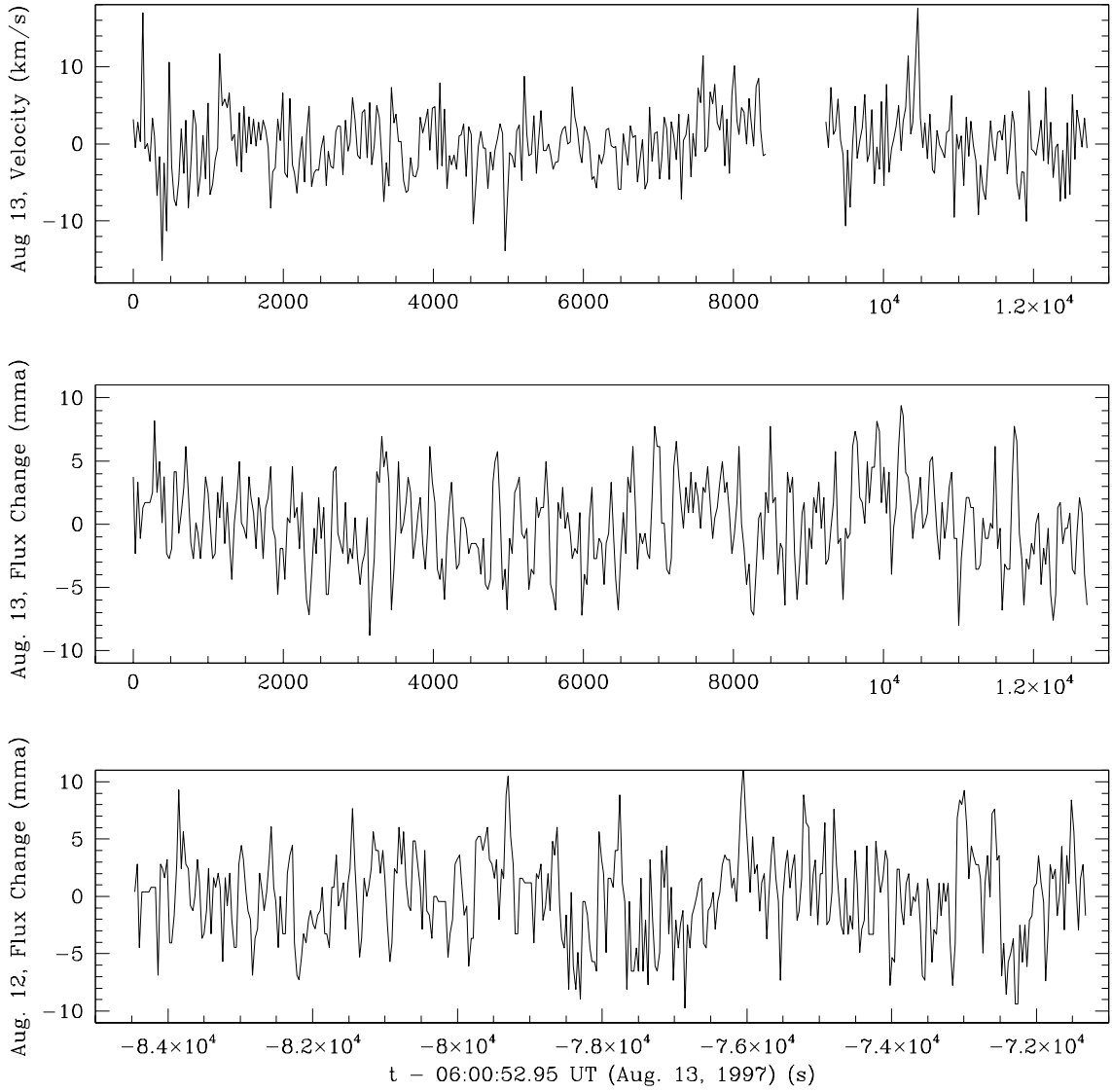}
\caption{The light curve for both nights and the velocities
measured on the second night. The portion of the velocity
curve near 9000~s was removed because it was obviously dominated 
by the motion of the star in the slit.}
\label{LC}
\end{figure}

\begin{figure}
\plotone{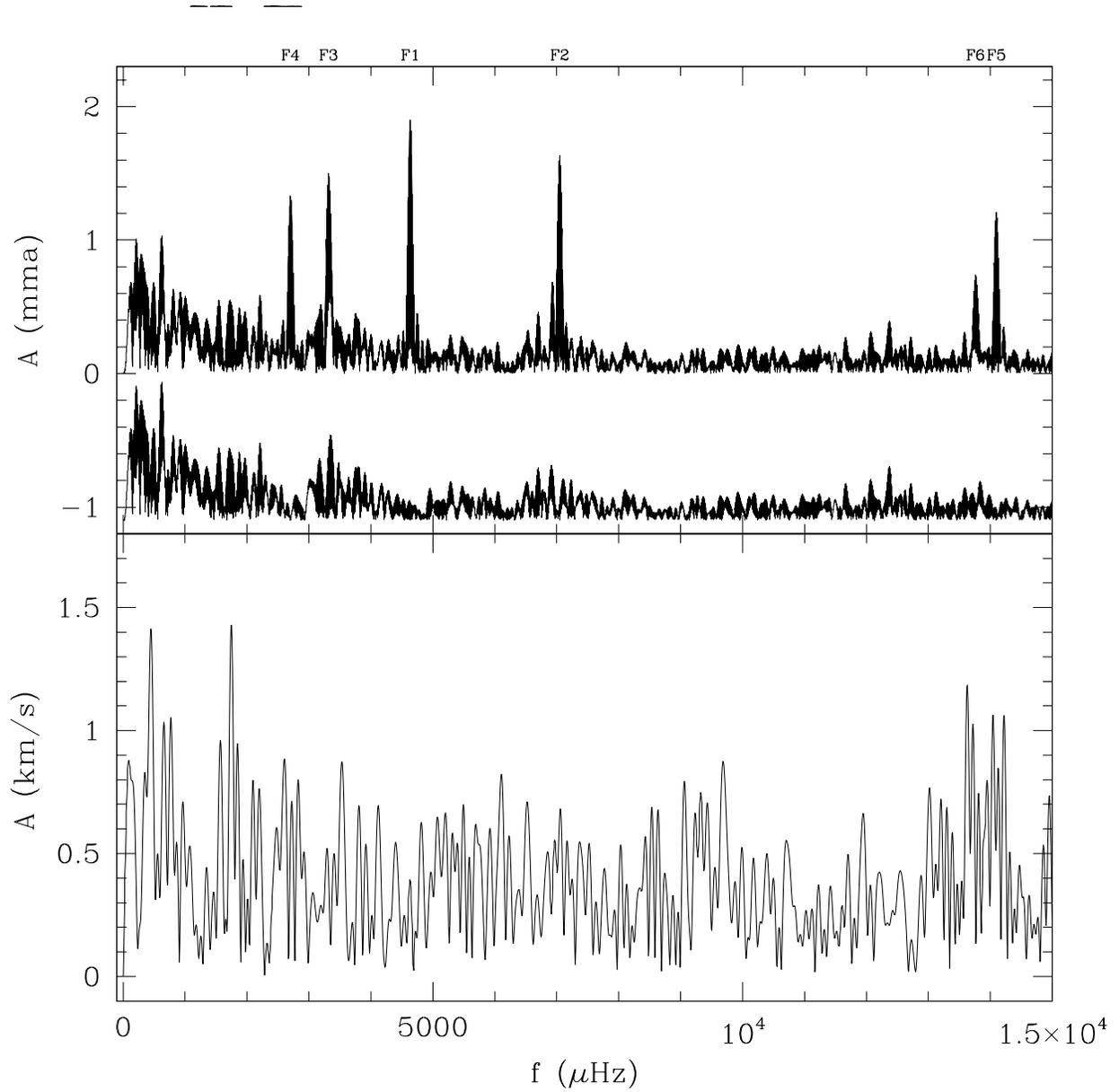}
\caption{The Fourier Transforms of the combined light curve of both nights 
and the velocity curve from the second night. The modes listed
in Table 1 are listed at the top of the light curve FT. An FT of the 
residuals after the six largest modes are removed is shown below the 
light curve FT.}
\label{FT}
\end{figure}

To create a light curve from the 7.4~hours of data, we
averaged the flux contained in the continuum between 5080 and 5420~\AA. 
We removed a 4th order polynomial, and
transformed the light curve to fractional variations (1~mma=0.1\%). 
Figure~\ref{LC} and~\ref{FT} show the light curve and its Fourier 
transform (FT). The six largest periodicities in this data, the
ones we will concern ourselves with here, have
all previously been observed on this star, most
recently during a 76 hour run with the Whole Earth Telescope 
\citep{WET}. Several other modes are present in \citet{WET},
including possible frequency splittings of our F2 and F3.
The mode labeled F2 is the perplexing 142~s mode noted 
in the introduction for its unusual behavior in the ultraviolet;
F5 is its harmonic.
 
The noise level of this FT, as determined from the square-root of the 
average power in the frequency range $8000-12000~\mu Hz$, 
is 0.1~mma \citep{HB86}. Each of the six modes listed in 
Table 1 are significantly above this noise level.  
The appearance of the light curves reveal that the nights were not perfectly
photometric; as such, we do not attempt to discuss any 
peaks in the FT not detected in previous observations. For our purposes, 
we are content with focusing on the six largest, previously observed modes.

Since we intended to measure the pulsation velocity variations,
we gauged how much the star moved in the slit during
the observations by measuring the location of the star
in the spatial direction at the 500th column of each image.  
A quick look at each curve, after a low order
trend was removed, revealed that on
the first night the motion of the star along the slit was 
a factor of ten larger than the second night.  
The standard deviation
of the motion of the star, translated into a velocity
for the same motion in the dispersive direction,
on the first night is 32~km~s$^{-1}$ 
while it is 2.5 km~s$^{-1}$ on the second.  
Measuring the velocity on each night by fitting the spectral lines
further confirmed this excessive motion on the first night.  
On the first night we used a B filter for the
the guide star, in a misguided attempt to remove the effects of
differential refraction. This made
the guide star too dim for the auto-guider to 
consistently maintain a lock on the star and resulted in
more random motion in the slit. No filter was used
on the guide star on the second night and the result was
much improved auto-guiding. Our analysis only 
includes the velocity measurements from the second night. 

To measure the wavelength shifts in each spectrum, 
we measured the central wavelength of the spectral
lines relative to the average spectrum.  We individually 
fitted the 5 Hydrogen lines (\Hb-H$_8$) with a 
Lorentzian and Gaussian function imposed
on a sloped continuum with the specfit routine 
of the STSDAS\footnote{STSDAS is a product of the Space Telescope
Science Institute, which is operated by AURA for NASA} package in IRAF.  
We required the Lorentzian and Gaussian to 
have the same central wavelength. The slope and flux of the
continuum, the flux in the Lorentzian and Gaussian, 
and the central wavelength varied to fit each spectrum.  
A fit to the average spectrum determined the initial conditions 
of each fit.   
We averaged together the velocities measured from each line, weighting 
each by the formal error of the fit.  We removed a low order trend
due to the differential refraction and flexure.  
Figure~\ref{LC} shows our velocity curve.

In contrast to the data on G~29-38 presented by \citet{VK00},
no obvious pulsations are present in the 
velocity FT of our data (see Figure~\ref{FT}).  
The few periodicities present just above the
noise level ($\sigma$=0.4 ~km~s$^{-1}$) are not present in the light curve's FT 
and are most likely due to motions of the star in the slit.
The analysis of this data can at most reveal an upper limit 
for the velocity associated with the flux variations ($\sim1.2$~km~s$^{-1}$).
Comparing our velocity FT to the similar LRIS data of G~29-38 
presented in \citet{VK00}, our velocity noise level 
is smaller than what they achieved.  If the velocity amplitudes
of \gone\ were as large as the velocities on G~29-38,
we would have been able to detect them in our data.

We performed a non-linear least squares fit to the six largest modes 
found in the flux curve and, for completeness, fitted the velocity curve 
of the second night at those same frequencies.  Table 1 shows the frequencies, 
amplitudes and phases of the six modes.  All phases 
are reported with time zero as August 13, 1997 at 6:00:52.95~UT.

\begin{deluxetable}{rccccc}
\tablenum{1}
\tablecolumns{7}
\tabletypesize{\small}
\tablewidth{0pt}
\tablecaption{The fitted modes of the light and velocity curves.
\tablenotemark{\dag}}
\tablehead{\colhead{mode} & \colhead{P(s)} & \colhead{$f$($\mu$Hz)} & 
\colhead{A$_f$(mma)} & \colhead{$\Phi_f$(deg)} & \colhead{A$_v$(km/s)}}
\startdata
F1 &215.7 &4634.9$\pm$.3 &1.9$\pm$.1 &77$\pm$5 & 0.37$\pm$.3\\
F2 &141.9 &7048.5$\pm$.3 &1.5$\pm$.1 &268$\pm$7& 0.64$\pm$.3 \\
F3 &301.6 &3315.8$\pm$.3 &1.5$\pm$.1 &281$\pm$7 &0.45$\pm$.3\\
F4 &370.2 &2701.1$\pm$.4 &1.3$\pm$.1 &202$\pm$8 &0.52$\pm$.3\\
F5\tablenotemark{*} &70.9 &14097.1$\pm$.4 &1.2$\pm$.1 &217$\pm$8 &0.55$\pm$.3\\
F6 &72.6 &13772.7$\pm$.7 &0.7$\pm$.1 &268$\pm$15 &0.17$\pm$.3\\
\enddata

\tablenotetext{*}{$F5=2*F2$}
\tablenotetext{\dag}{The errors reflect the formal errors of the fit.
The reported velocity amplitudes are not significant detections.}
\end{deluxetable}

\subsection{Chromatic Amplitudes}
\begin{figure}
\epsscale{0.65}
\plotone{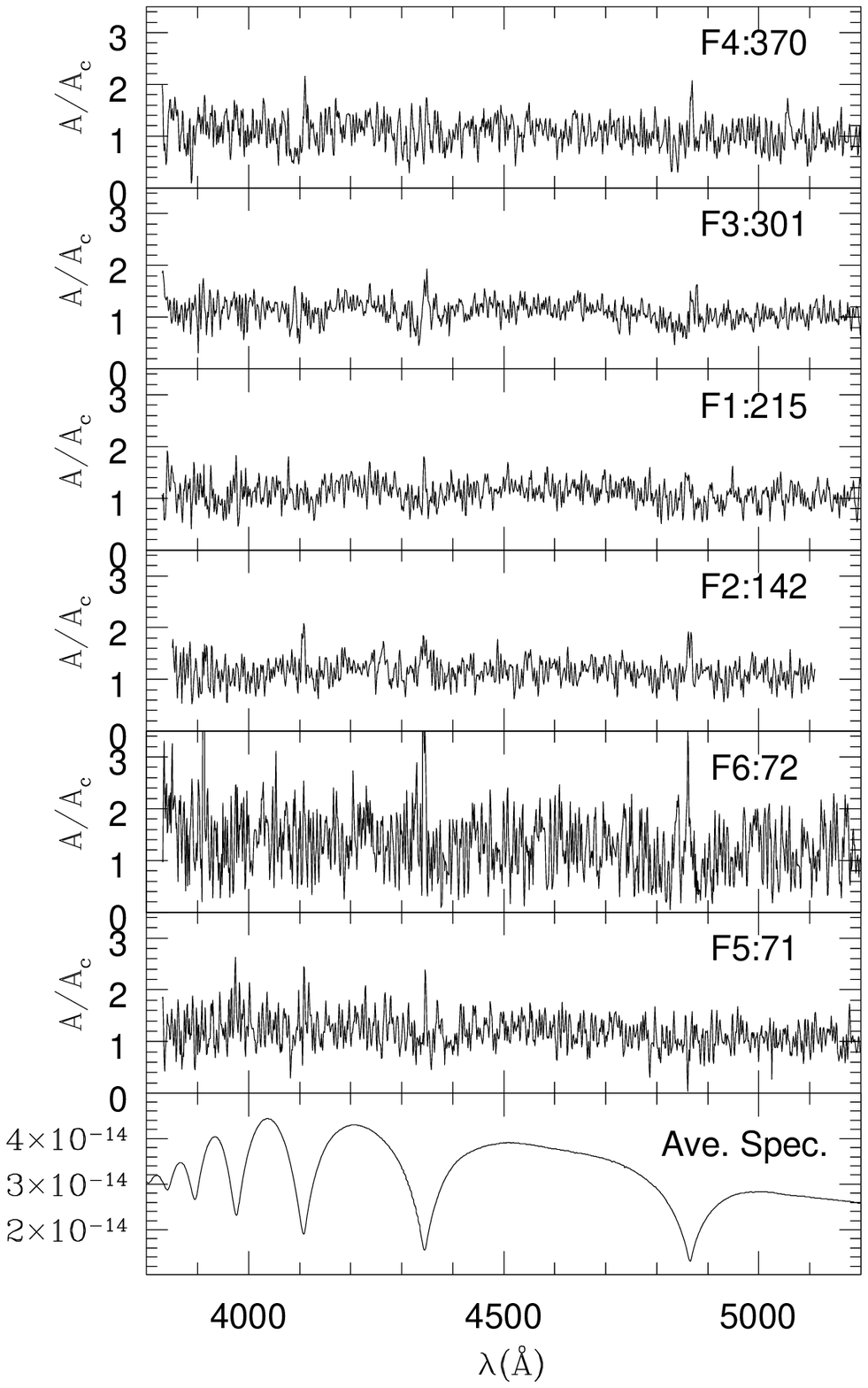}
\caption{Chromatic amplitudes for each of the modes in Table 1 determined
directly from the data. Each plots the fractional flux change as a 
function of wavelength normalized by the continuum amplitudes in Table 1.}
\label{dataca}
\epsscale{1}
\end{figure}

  In hopes of determining the spherical degree
of each of the observed modes, 
we created a light curve at each wavelength and measured the 
amplitude from the FT for the six modes in Table 1.  
The plot of amplitude versus wavelength, 
named chromatic amplitudes by \citet{VK00},
were normalized by the amplitudes measured from 
the continuum light curve. Figure~\ref{dataca} shows the
disappointing chromatic amplitudes determined
directly from the data. Except for the amplitude peak
that occurs at the center of some spectral lines, 
this analysis revealed nothing useful.

  The reason for the poor chromatic amplitudes is the low signal-to-noise
ratio in any pixel ($\sim 1.23$~\AA). 
To circumvent this problem we attempted to smooth the data, 
effectively increasing the wavelength coverage of each element. 
The application of a boxcar smooth to our spectra 
did decrease the noise in each element. 
However, when we sufficiently reduced the noise by
averaging together a larger number of pixels, 
the wavelength coverage of any one bin became so large
that the spectral lines were washed-out.
In spite of these discouraging results, we endeavored to find
another method to measure the chromatic amplitudes.
We discovered that when we imposed information about the 
shape of the spectral line derived from the high signal-to-noise
average spectrum, we were able to improve
the appearance of the amplitude variations.   
Essentially, we fitted the average spectrum to establish
an expected shape of the spectra and then fitted each 
individual spectrum, allowing variations only in a few free parameters.  
We then used these fits to measure the chromatic amplitudes instead of the
original data. Constraining the variations of the line shape in this way
has greatly reduced noise while still maintaining information about the
actual line profile variations.
Since this is a new method for analyzing
this type of data, we continue by providing more details.

\begin{figure}
\epsscale{0.75}
\plotone{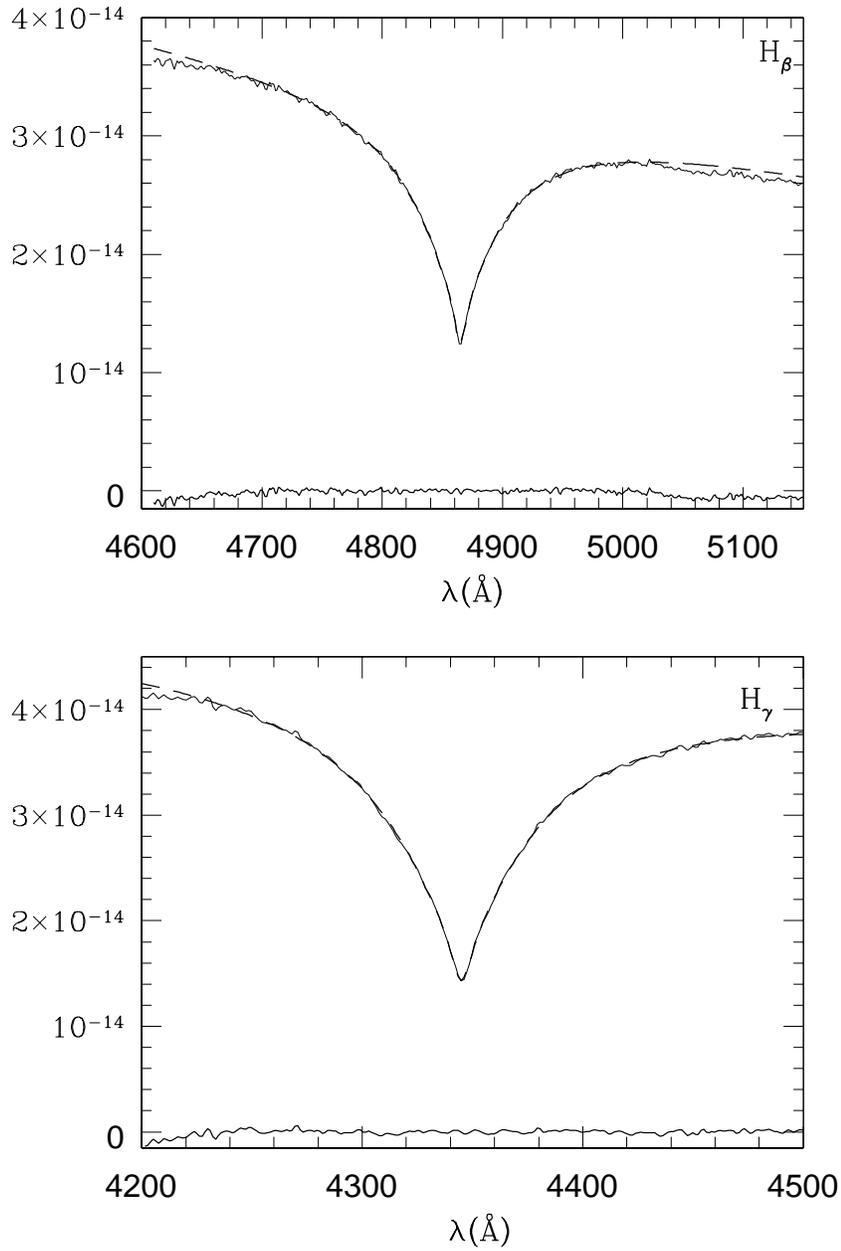}
\caption{The solid curve shows an individual spectrum while the dashed line is
the fit to that spectrum. Top panel is \Hb\ and the bottom panel is \Hg. 
The residuals are plotted at the bottom of each panel.}
\label{fitsp}
\end{figure}

We applied the same technique to both the \Hb\ and \Hg\ lines.
First, we measured the spectral shifts of the line itself by fitting
the line as described in \S2.2.  We then removed those spectral shifts 
because the velocities due to the motion of the star were too small to 
observe and the spectral 
shifts would not add useful information to the results. Additionally, 
removing the velocities ultimately makes comparison with the models simpler. 
Next, we created an average spectrum from the deshifted spectra and used its fit 
as the initial conditions for the fit to the individual spectra.  
The same parameters were used here as for measuring the velocities, 
a Gaussian and Lorentzian applied to a flat continuum.  
As we fitted each spectrum, 
we only allowed four parameters to vary: the continuum flux, the continuum slope,
and the flux in the Gaussian and Lorentzian. 
Since we removed the spectral shifts, the central wavelength 
of the fit is fixed to that of the average spectrum.   
We show a typical example of a fit to an individual spectrum and the residual
in Figure~\ref{fitsp}.
We fitted across the region 4220-4470~\AA\ for \Hg\ and the region 4760-4960~\AA\ 
for \Hb.  

\begin{figure}[!ht]
\epsscale{0.53}
\plotone{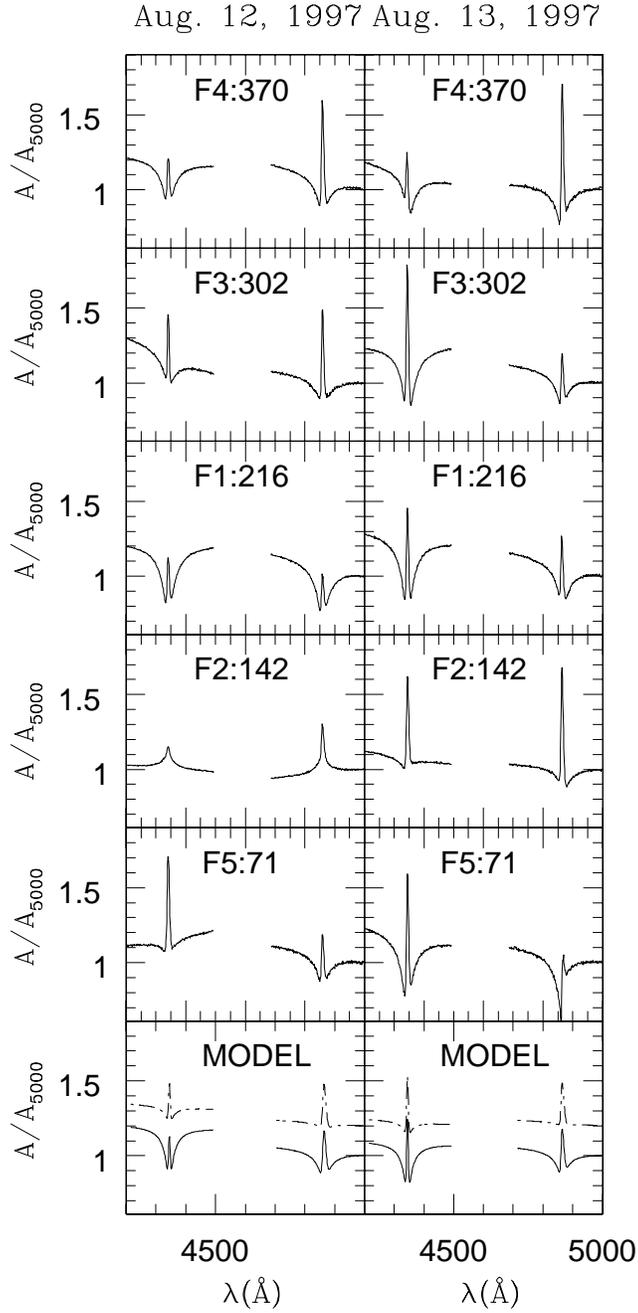}
\caption{Chromatic amplitudes of the \Hb\ and \Hg\ lines created
from the fits to the spectral lines for both nights of data.
Each has been normalized at 5000~\AA. The bottom plot contains model
chromatic amplitudes of \el=1 (solid) and \el=4 (dash) created
with fake spectra gaussian smoothed according to the resolution
of each night (see \S\ref{fake}). For clarity, the normalized 
amplitude of the \el=4 model has been offset by 0.2.}
\label{cafit}
\end{figure}

\begin{figure}[!th]
\epsscale{1}
\plotone{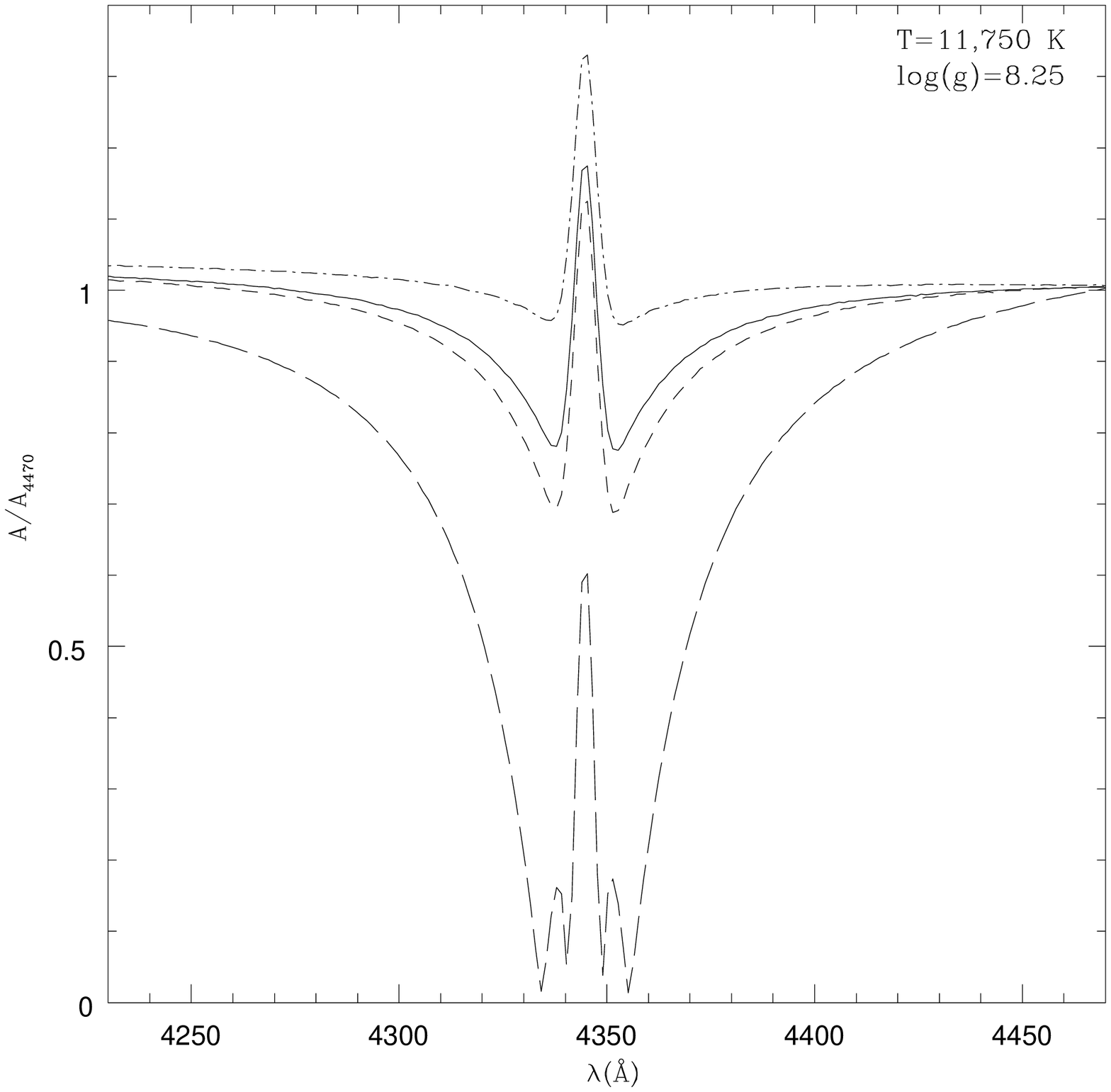}
\caption{Chromatic Amplitudes of \Hg\ created from fitting model spectra.
Each spectrum was created by scaling model spectrum 
by the expected model chromatic amplitudes in Figure 1 and using
a fit to re-create the chromatic amplitudes.  Each have been normalized
to 4470~\AA\ and shifted in wavelength to align with our spectra.
The first four spherical degrees are shown as solid, short-dashed, long-dashed,
and dot-dashed in order of increasing \el.}
\label{fitmodel}
\end{figure}

We then created chromatic amplitudes from the fits to each
spectra instead of directly from the data.
For each wavelength we created a light curve of fractional changes in flux 
and took a Fourier transform at each wavelength.  We determined the 
amplitude of the FT at each of the six modes listed in Table 1
and normalized the curves by the amplitudes measured
at 5000 \AA.
Figure~\ref{cafit} has the plot of these chromatic amplitudes
for the \Hb\ and \Hg\ lines on each day for the four
largest modes. The noise level of the FT at each 
individual wavelength is approximately 0.2~mma. As a result,
the low amplitude modes, F5 and F6, are not significantly above
the noise and show sporadic results.

\subsection{Spectral Fit Chromatic Amplitudes}

The chromatic amplitudes are a
way of examining how the shape of the spectral line changes
during a pulsation cycle. This depends on how
the flux variations are distributed across the surface of
the star, which change with \el. In Figure~\ref{model}
we showed model calculations for how the chromatic amplitudes 
differ for values of \el=1 to 4.

We are encouraged that the chromatic amplitudes 
in Figure~\ref{cafit} have
the same basic features as the model chromatic amplitudes.
Since we did not fit the continuum to create the chromatic amplitudes,
we do not expect the region outside of the fitted wavelengths to 
closely follow the models. Hence, we do not
see the arches between the spectral lines apparent in the models.
Also, the seeing on the first night was larger than the second,
possibly washing out more of the information concerning the changing
line shape. This may account for the differences between 
the chromatic amplitudes on different nights. 

The drastic improvement in the appearance of
chromatic amplitudes seen in Figure~\ref{cafit}
is very encouraging. However, the danger in using
spectral fits instead of the original
spectra is a possible introduction of systematic effects that
could alter the final results.  We are making an assumption
about the appearance of the spectral line that might not be 
entirely accurate.

\label{fake} To test how our technique alters the appearance of
the chromatic amplitudes, we created a time series of 410 fake spectra. 
We introduced the line profile variations to the spectra by adding
a product of the spectrum, the model chromatic amplitude,
and $\cos(\omega t+ \phi)$.  We introduced a strong 
signal for modes from \el=1 to \el=4, convolved each spectrum
with a gaussian to emulate the resolution of our data (3.9~\AA), 
and then produced chromatic amplitudes in the same way described above by
fitting each spectrum with a Lorentzian plus a Gaussian.  
The chromatic amplitudes created by this experiment 
(Figure~\ref{cafit} and \ref{fitmodel})
show a few differences from the original model chromatic amplitudes
in Figure~\ref{model}.  The new model chromatic amplitudes
show a narrower peak and shallower dips on either side, 
making their appearance similar to our chromatic
amplitudes in Figure~\ref{cafit}. 

\begin{figure}[!ht]
\epsscale{1}
\plotone{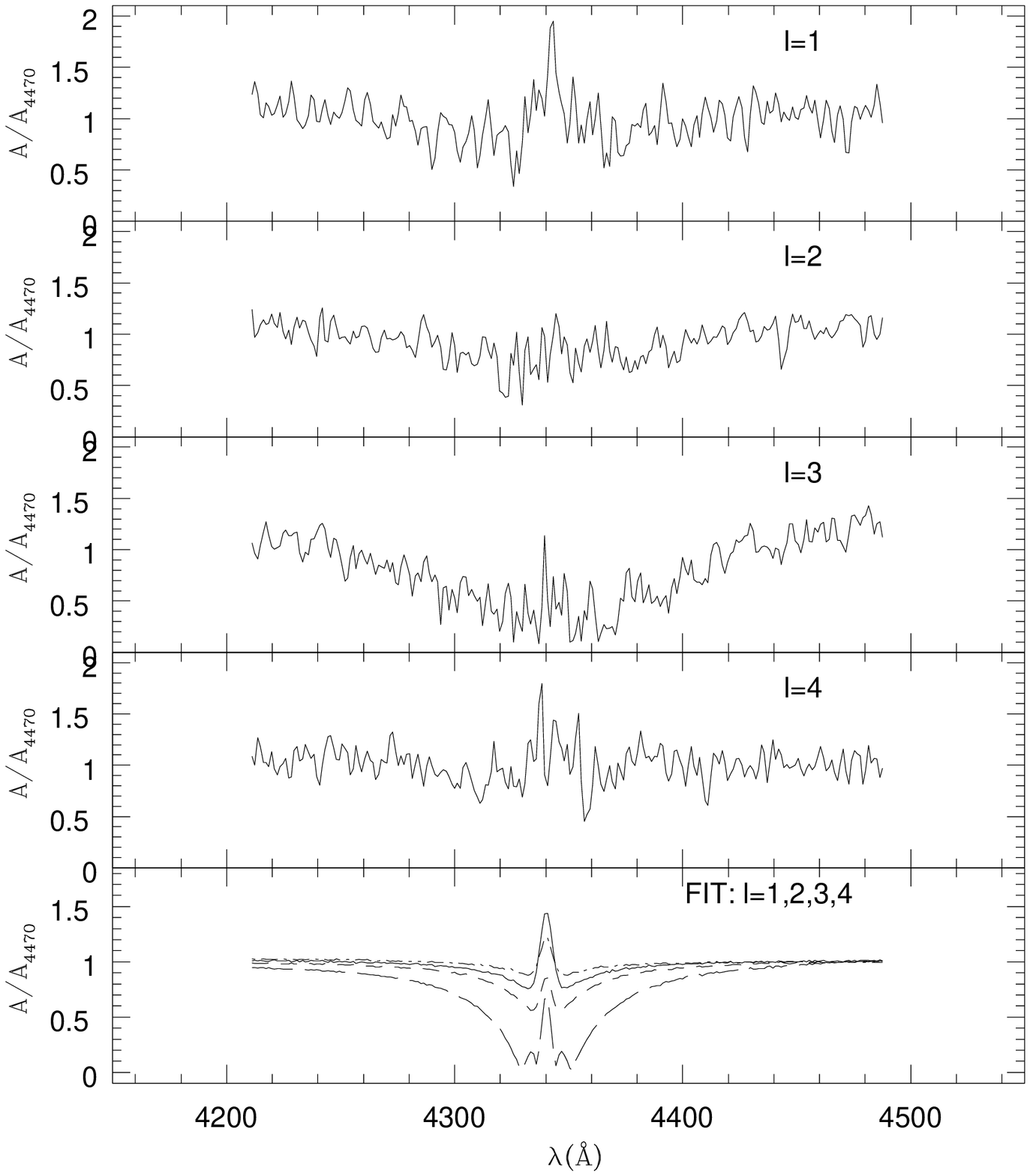}
\caption{Chromatic amplitudes created from noisy simulated spectra.
The first four show the chromatic amplitudes as measured
directly from the fake spectra. The bottom panel shows 
all four chromatic amplitudes created from the fits 
to those spectra. The lines are solid, short-dashed,
long-dashed, and dot-dashed in order of increasing \el.}
\label{fakecrap}
\end{figure}

To further test the validity of this technique 
we added random noise to the same series of fake spectra, 
giving them the signal-to-noise of our spectra. 
We used the model chromatic amplitudes to introduce 
pulsations of different spherical degrees 
with amplitudes near 2 mma at 4470~\AA. 
We then measured the
amplitudes at each wavelength both directly from the noisy fake spectra
and the fits to the spectra.  The results of this experiment
can be found in Figure~\ref{fakecrap}. The raw chromatic amplitudes
resemble those from our data in Figure~\ref{dataca}. In spite of their poor appearence,
application of the spectral fitting technique to the noisy fake spectra
successfully recovered the shape of the model chromatic amplitudes.  
From this experiment we have shown that the addition 
of information about the shape of the spectrum can overcome
a series of spectra with a signal-to-noise too poor to create
meaningful chromatic amplitudes directly from the data.

We performed the same experiment for different periods and different
amounts of noise in order to see how they could affect 
the appearance of the chromatic amplitudes.
Even with moderately noisier spectra, the basic distinctions 
between the different spherical degrees remain. 
The largest variation from the models occurs at the 
central peak; noise can
both increase and decrease the height of this peak.
The line center is described by the
Gaussian portion of the fit and is dominated by
only a few points, so we might expect 
noise to have a greater effect on the 
central region of the line.

In the fits shown in Figure~\ref{cafit},
the chromatic amplitudes of F1, F3, and F4 all
appear to have the familiar 'w' shape of the \el =1 or 2 modes.
They mostly show no significant asymmetries, expected since
we removed the velocities prior to the fits.
The chromatic amplitudes of F2, the 142~s mode, measured for  
\Hb\ and \Hg\ on both nights shows a distinctly different
shape. This mode has almost no drop on either side of the central peak.
When compared to the model chromatic amplitudes in Figures~\ref{model}
and~\ref{fitmodel}, it most resembles the characteristics
of the \el=4 model. All modes identified
on DAVs thus far have had \el$\le 2$, expected
since higher spherical degrees suffer from increased 
geometric cancellation.

\begin{figure}[ht]
\epsscale{1.0}
\plotone{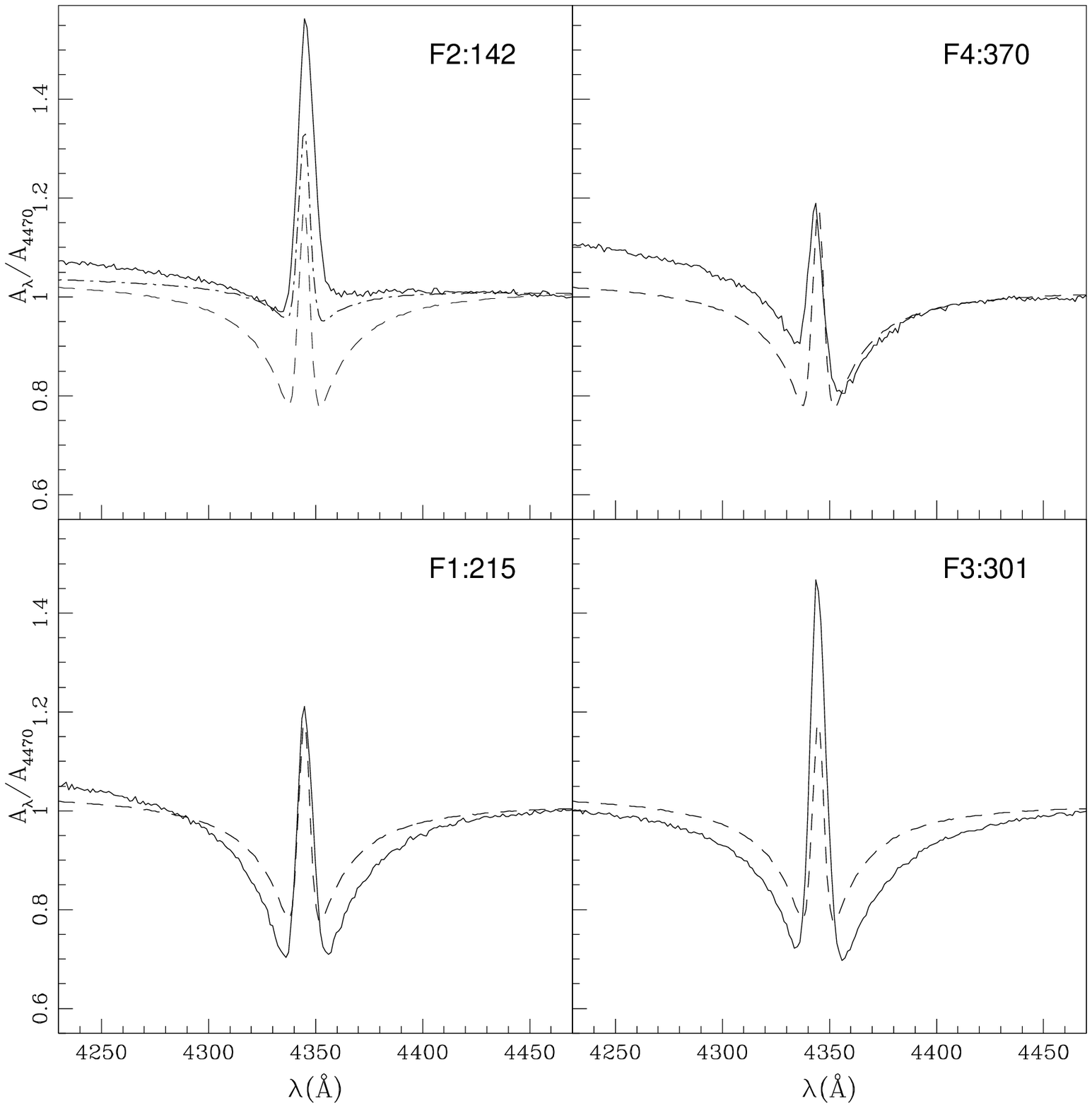}
\caption{Comparison of the fitted model chromatic amplitudes
to the fitted observed chromatic amplitudes. 
The solid line is the observed chromatic amplitude from the 
gamma line on the second night and 
the dashed line is the \el=1 model. For the 142~s mode the dot-dash 
line is \el=4.
Models were calculated using a mean temperature of 11,750~K and log(g)=8.25.
Each curve has been normalized at 4470~\AA.}
\label{mfmodel}
\end{figure}

For direct comparison we show the chromatic amplitudes
for the second night of \Hg\ of the four largest modes along with
the model chromatic amplitudes created by fitting the simulated spectra 
(Figure~\ref{mfmodel}). In order to make an accurate comparison, the
simulated spectra were shifted in wavelength to agree with the central
wavelength of our average spectrum.  F1, F3 and F4 all
show similarities with an \el=1 mode while F2 resembles
an \el=4. The discrepancy between the model and our chromatic amplitudes
lies mostly in the central peak.  The bottom of the spectral line has
a smaller signal-to-noise and as we discussed earlier, noise appears
to mostly affect the size of that central peak.  Regardless, 
the differences between the \el=1 and \el=4 remain 
and can clearly be seen in the plot of F2 in Figure~\ref{mfmodel}.

As we shall see, the idea of an \el=4 is not as outlandish
as it first seems.  An \el=4 mode, as suggested by the chromatic 
amplitudes, can also explain the unusual behavior
in the UV, the large harmonic mode, and the low amplitudes
of \gone. In the next section, we will discuss the previously
published observations in light of this new hypothesis.

\section{The \el\ $= 4$ Hypothesis}

  The addition of optical time-resolved
spectroscopy to the observations of \gone\ has
led us to present a hypothesis that potentially 
explains all the observations of this star's pulsations. 
The frequency spectra of simple, blue-edge DAVs are commonly
dominated by one mode.
In this section, we consider our F2, the 142~s mode, 
to be the dominant pulsation mode of this star and 
to have a spherical degree of four. The remaining
modes are either weakly driven \el=1,2 modes or combination
modes.  We will now consider to what extent this hypothesis
can account for previous observations of \gone.

\subsection{UV observations with HST}
In the UV observations presented by \citet{Ke00}, 
the relative amplitudes of the modes on \gone\ significantly 
increase at shorter wavelengths except for the 142~s mode.
The amplitude rise is a result of both the increased effect
of temperature on flux and the increased limb darkening at these
shorter wavelengths.
The 142~s mode is perplexing because it does not match the other modes
or the models of \el$\le2$. 
Considering modes of higher \el, we note that 
in Figure~\ref{model} the \el=4 mode does not 
increase in amplitude like the lower order spherical degrees. 
The negative amplitudes shown in the model 
would be measured with a positive amplitude and a 180 degree phase shift.
We examined the UV data presented in \citet{Ke00} to judge 
the agreement of the phases and amplitudes of the 142~s mode with
the \el=4 model.
 
\begin{figure}[!ht]
\epsscale{1}
\plotone{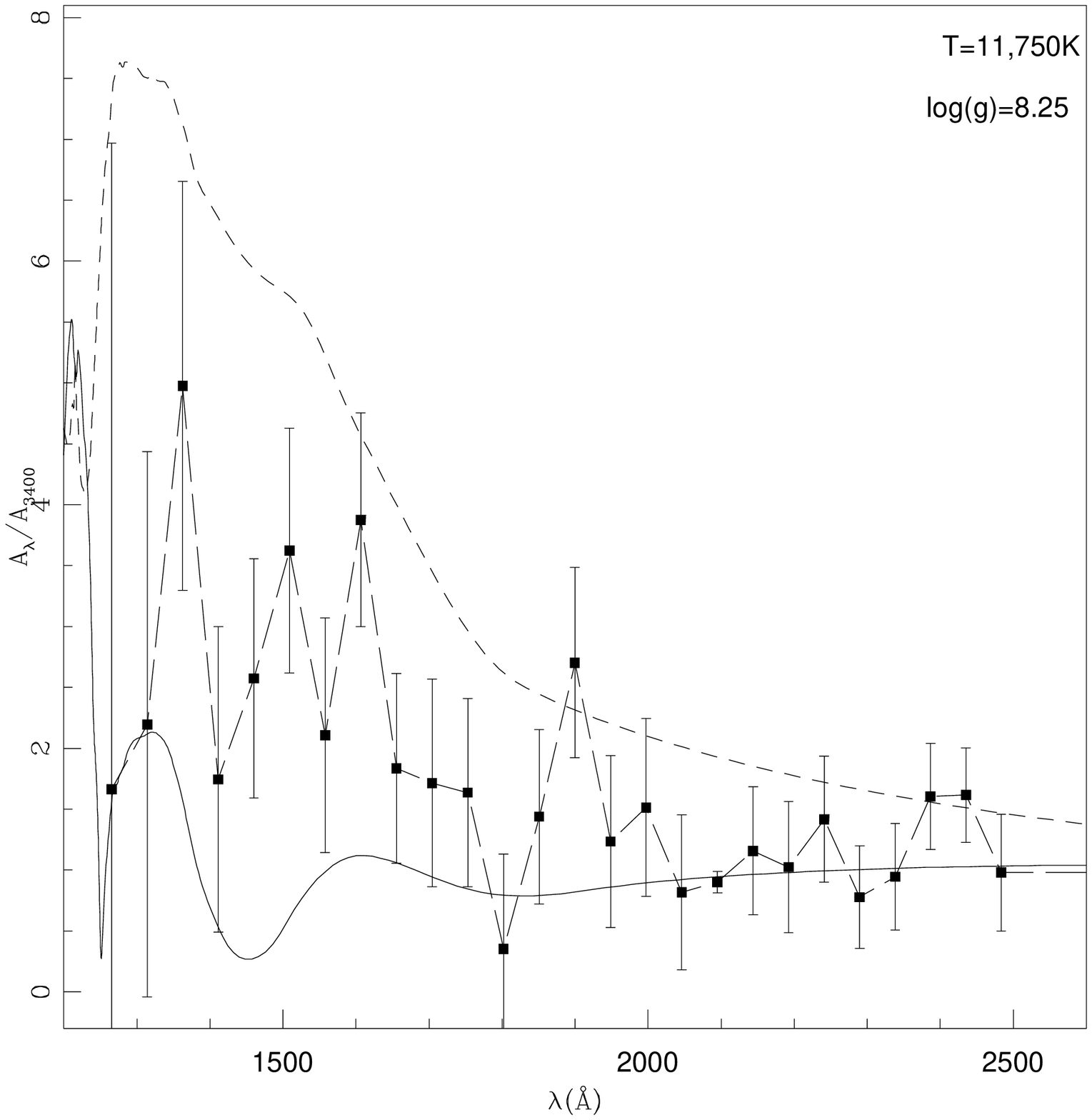}
\caption{Amplitude variations in the UV. 
Absolute value of the \el=1 (short dashed) and \el=4 (solid) 
model plotted with the measured amplitudes
of the 142~s (F2) (long dashed). The 
observed UV amplitudes are from the HST data \citep{Ke00}. }
\label{modeluv}
\end{figure}

To obtain the amplitudes and phases at each wavelength of the HST data
we fit the zero-order light curve ($\sim$3400~\AA) 
with the seven dominant modes to find the frequencies, amplitudes and phases.  
We then fixed the frequencies, fit each wavelength bin with all seven modes  
and normalized the amplitudes and phases by the zeroth
order fit. The relative amplitudes are presented in Figure~\ref{modeluv}
along with the \el=1 and \el=4 models.  As mentioned
before, the measured amplitudes do not rise as quickly as the \el=1 model
and are more consistent with the \el=4 model.

\begin{figure}[!b]
\epsscale{1}
\plotone{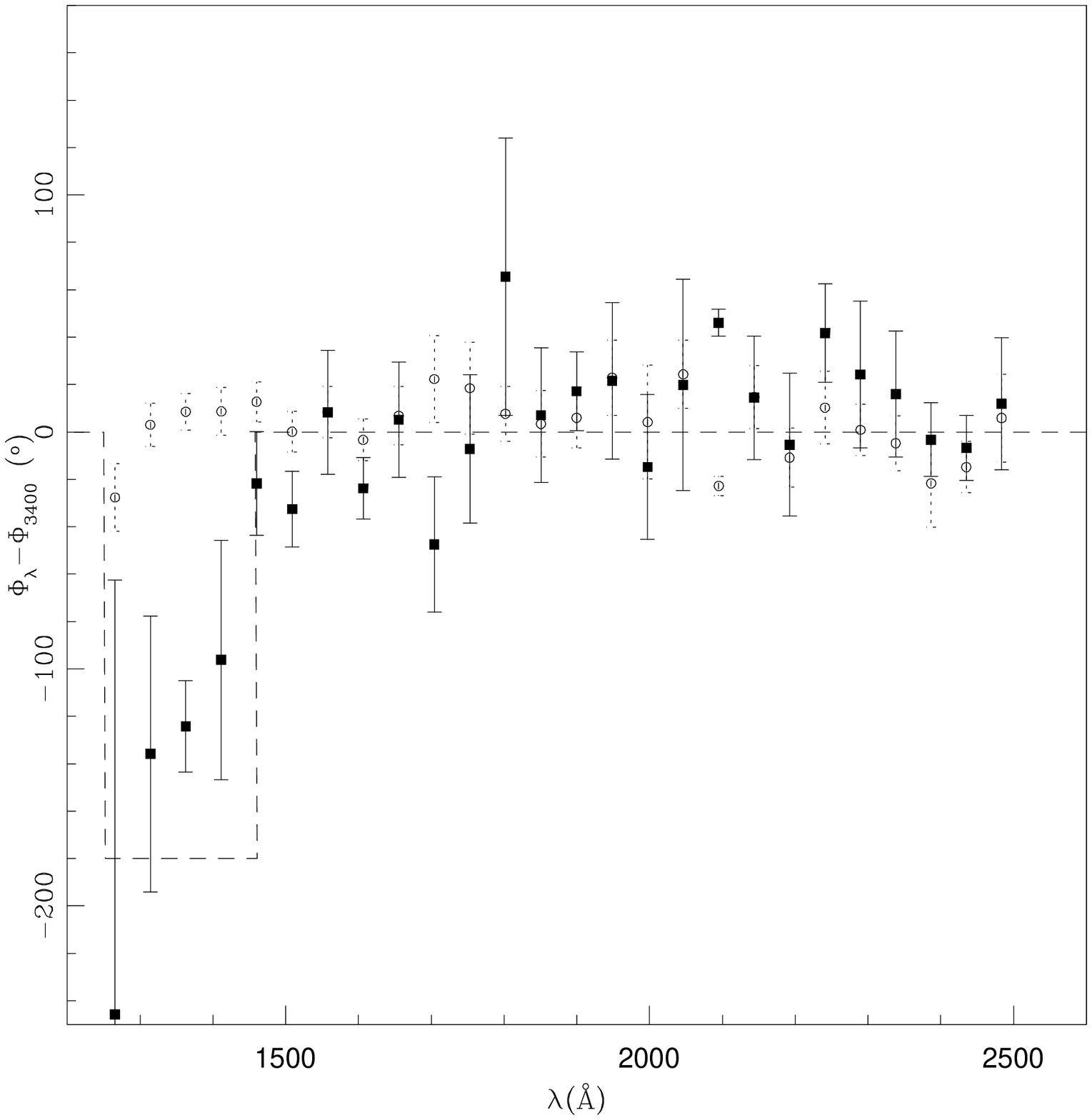}
\caption{The phases for the 142~s (F2) (squares) and the 71~s (F5) (circles) 
modes relative to the phase measured at 3400~\AA.  The error bars
reflect a reduced chi-square of one for each wavelength bin. The dashed line
represents the phases expected for an \el=4 model. The \el=1,2,3 models
would show no phase change.}
\label{phase}
\end{figure}

Figure~\ref{phase} shows the phase
at each wavelength for the 142~s mode and its harmonic (71~s).  
While the harmonic shows no phase reversal, at
wavelengths less than 1500~\AA, our proposed \el=4 mode shows
three points with a phase change greater than 100$\degr$.  
Though the change in phase of those points is less than 180$\degr$,
they occur where the \el=4 model shows a phase
change. 
We conclude that while these measured phases do not prove 
the \el=4 hypothesis, they also do not exclude it.

The minor inconsistencies between the data and the \el=4 model
could be in part due to third order combinations with the same frequency
as F2 (i.e. F2=2F2-F2). Though small in amplitude, 
they would have low \el\ components, increasing in amplitude
and having no phase change at UV wavelengths.  The added
amplitude and different phase of this mode could interfere
with F2 and create the small discrepancies with the \el=4 model.

We note that the 142~s mode is not the
only mode that shows a curious UV behavior. 
Observations of the
DAV, G~117-B15A \citep{K03}, reveals a mode that
does not increase in the UV, but no other evidence
suggests that this mode should be an \el=4.
Also, unlike our observations, the same mode on G~117-B15A shows
an unusual trend at the blue end of the optical wavelengths.
Apparently, this failure to rise in the UV is not an 
exclusive signature of a higher degree mode. 

Some of the inconsistencies
could result because it might not be entirely 
appropriate to use the linear models for comparison
with the measured wavelength dependence of the amplitudes and phase.  
\citet{IK} performed non-linear simulations and showed that
the size of the amplitude and inclination can play a role in 
how the relative amplitudes and phases
vary with wavelength in both the UV and optical wavelengths.  
Their study concurs that a failure to rise
in the UV is consistent with \el$>2$.  We conclude that 
though the UV data is not a perfect match to the \el=4 hypothesis, 
it is a better fit to the observations than any other available choice.

\subsection{The Harmonic Mode}
  We consider the implications of the \el=4 hypothesis on the harmonic of the
142~s mode. F5, the 71~s mode, is exactly twice the frequency of the proposed
\el=4 mode.  \citet{WET} found that our F2 is a combination of F6 (72 s) and
a mode at 148~s. In our hypothesis F5 (and F6 or 148~s) is not a real mode;
it is the result of nonlinear effects most likely caused by the surface
convection zone \citep{Wu01, IK}. Large combination modes are uncommon on
blue-edge DAVs like \gone, and unexpected for low amplitude pulsations.  If
F2 is indeed an \el=4 mode, its actual amplitude is much larger than
observed, removing this inconsistency with the theories.

\citet{Wu01} has provided an analytic description of how
a harmonic mode is distributed on the surface of
the DAV, and concluded that a harmonic mode appears 
as the square of the parent mode's distribution.  
From this theory, only two scenarios
can create a harmonic similar in size to its parent.
First, a large inclination of the star, as proposed by \citet{TC03},
could provide large cancellation of the parent mode but leaves
the harmonic unaffected. However, as we shall discuss 
in \S\ref{other}, this theory is unable to explain the 
UV characteristics of \gone.
Second, if the parent mode has a larger spherical degree,
it will suffer from cancellation over the surface of
the star while the harmonic mode, appearing as the
square of the parent, does not.
Though the total power of the harmonic mode is less than the power of its
parent, the severe cancellation of the parent mode can make
the harmonic mode appear larger in the observed light curve.
Having the 142~s mode be
an \el=4 mode is consistent with this picture; the 142~s mode is
innately canceled while its harmonic at 71~s is not. 

\begin{figure}[!ht]
\plotone{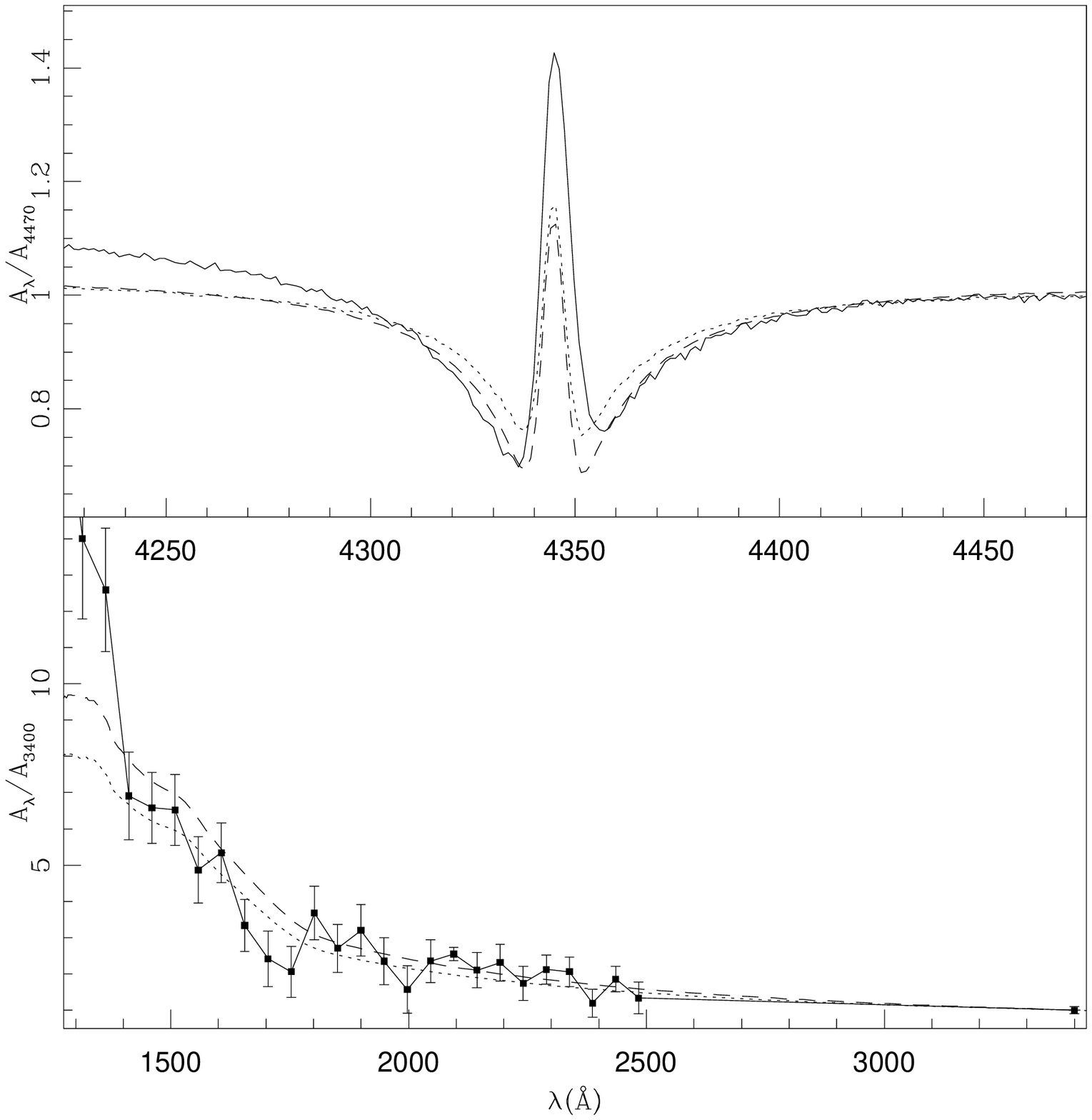}
\epsscale{1}
\caption{Comparison of models (Dotted is \el=2 and dashed is (\el=4, $m=0$)$^2$) 
to the optical chromatic amplitudes and UV data (squares) \citep{Ke00}
for F5, the 71~s mode.  The optical model chromatic amplitudes are 
created from simulated spectra and treated in the same way as the
data.  For both figures we used the model corresponding to 
T=11,750 and log(g)=8.25.}
\label{modelsq}
\end{figure}

If the chromatic amplitudes are modeled in the 
same way as Figures~\ref{model} and~\ref{mfmodel} for the surface
distribution of a harmonic, we notice a resemblance to
an \el=2 mode. This similarity is not surprising;
the square of an \el=4 spherical harmonic has
most of its flux changes at the pole and suffers from
no cancellation over the surface of the star just like 
a lower \el\ mode. Figure~\ref{modelsq}
compares the expected variations of amplitude with wavelength for
an (\el=4, $m=0$)$^2$ distribution to 
both the HST data and our \Hg\ chromatic amplitude of
the 71~s mode from the second night. Both curves
show agreement with the models.

The \el=4 scenario implies that the parent mode
has a low value of $m$. 
A harmonic created by an \el=4, $m\ge2$ mode will
resemble an \el$\ge 4$ mode.  Only combinations created
by \el=4, $|m|\le$1 modes will have a large amplitude,
increase in the UV, and have chromatic amplitudes similar to
\el$\le$2. Similarly, any azimuthally split modes, as the possible
one observed by \citet{WET}, would have harmonics with large
\el\ characters and small amplitudes.

\subsection{Optical Amplitudes}
If the 142~s is the dominant \el=4 mode, then this might
explain the low amplitudes of this star's pulsations.
The measured amplitude of an \el=4
mode would be limited by cancellation as 
the observed flux variations result from a sum across
the visible surface of the star.
To test this, we calculated how
much brighter an \el=4, $m$=0 mode would look if it appeared as an \el=1, $m$=0 mode.
\label{scale} To determine this factor we integrated both modes over
the visible surface of a star seen at its pole 
using limb darkened models
(T=11,750, log(g)=8.25) and summing the flux change over
the visible wavelengths (4000-5500~\AA).
We concluded that an \el=4 mode would be observed approximately 13 times
smaller than an \el=1.  So, if \gone\ is dominated by an \el=4 mode,
its amplitude has been reduced by a factor of thirteen, meaning
that our F2 would be $\sim19.5$~mma if it were an \el=1 mode.
Hot DAV stars have been observed with amplitudes as high as 23~mma
(see \citet{Ke91} and \citet{C94}).

\begin{figure}[!th]
\epsscale{1.0}
\plotone{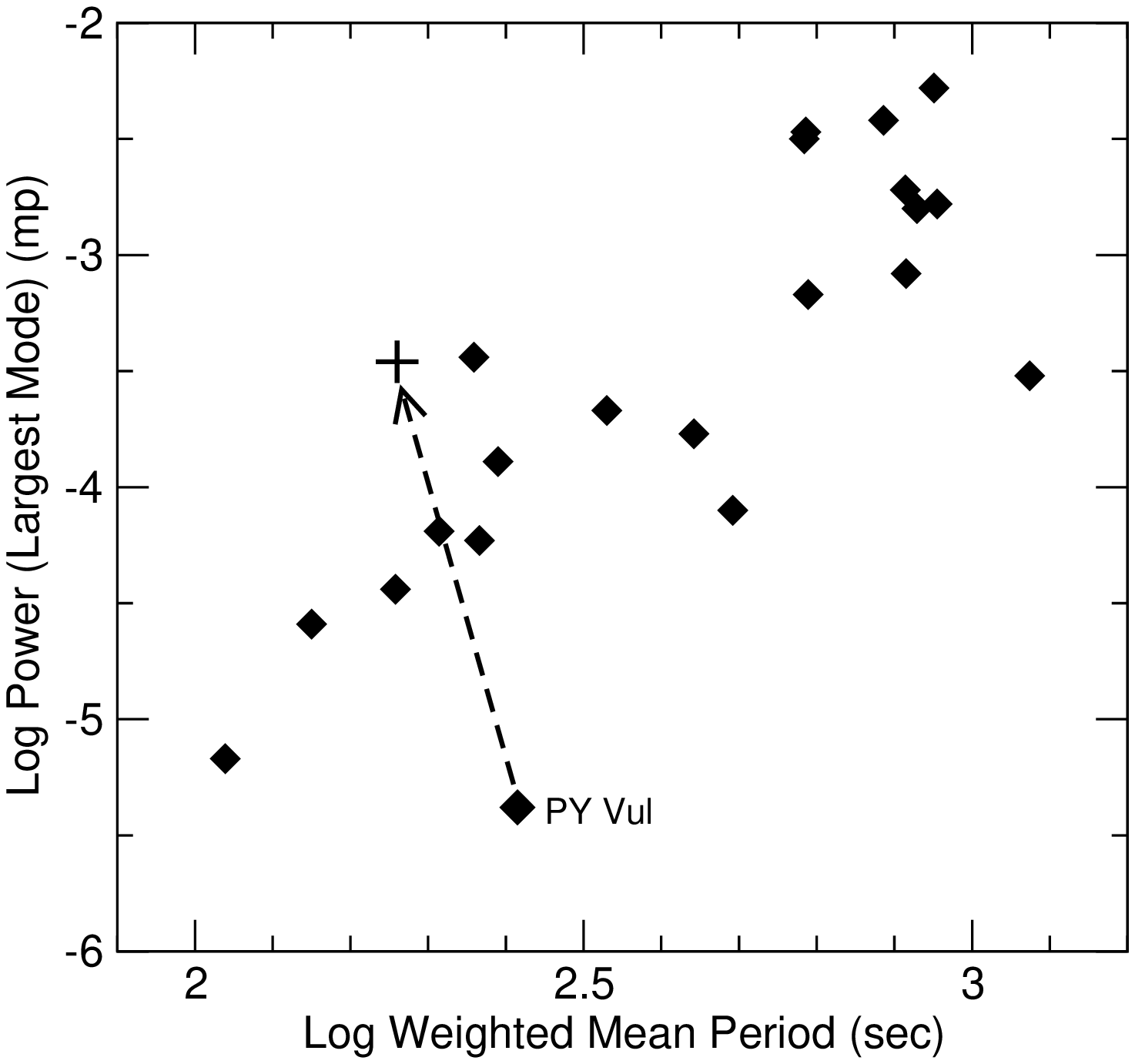}
\caption{The power of the largest mode plotted against the weighted mean 
period of the DAVs. The diamonds represent the 21 DAVs studied by Clemens (1994). 
The observed location of \gone\ is the diamond labeled at the bottom of the plot. 
We propose that the reason \gone\ has a lower power than the other DAVs is
that it is dominated by an \el=4. The other DAVs are expected to be
dominated by \el=1 modes (Clemens 1994).  Due to its inherent cancellation,
an \el=4 mode will appear 13 times smaller than an \el=1
(see \S\ref{scale}).  We apply this scaling factor to the 142~s mode of 
\gone\ and adjust for the new weighted mean period (the plus sign).   
A \gone\ showing an \el=1 mode instead of an \el=4 mode agrees with the
trend of the other DAVs, demonstrating that the \el=4 
hypothesis correctly accounts for the low amplitudes of \gone.}
\label{davs}
\end{figure}

Assuming that the other DAVs are generally dominated by
\el=1 modes as deduced by \citet{C94}, we compared 
\gone\ to the class of DAVs after applying
this scaling factor.  To illustrate the results, we reproduced 
the plot of weighted mean period verses power from \citet{C94} 
and added the location of an \el=1 dominated \gone\ (Figure~\ref{davs}). 
While the actual location of \gone\ has a low amplitude, 
placing it below the trend of the other DAVs, a \gone\ 
that does not suffer the geometric cancellation
of an \el=4 mode would be located at the plus sign in Figure~\ref{davs}.
Assuming \gone\ is dominated by an \el=4 mode brings the amplitude
of this star into line with the other stars of its class.

The \el=4 hypothesis satisfyingly explains all the above characteristics
of the pulsation spectrum of \gone.  Our only hesitation is the failure
to see 8 azimuthally split modes associated with the \el=4 mode. However,
we also do not see triplets for the other modes.
Either the star does not excite higher $m$ modes, or is rotating so slowly
that we have failed to resolve them.

\section{Other Proposed Scenarios}
\label{other}
Other suggestions have been made to 
explain the character of the pulsations of \gone. 
After a preliminary analysis of the same data presented 
here, \citet{TC03} proposed that a large inclination could
account for the low mode amplitudes. In this scenario, the 142~s
mode is the first harmonic of a mostly cancelled mode found near 285~s, while
the 71~s mode is the third harmonic. This hypothesis
did explain the failure of the 142~s mode to rise in the UV. 
However, it failed to explain the behaviour of the 71~s mode in the UV. 
Furthermore, the 285~s parent mode required by this scenario was
only marginally detected on the first night of the LRIS observations,
and the extensive coverage of the WET run on \gone\ \citep{WET} 
failed to find any significant signal near 285~s. This requirement
of almost total cancellation imposed by the WET data requires
a very extreme inclination. 
To achieve the ratio between the amplitudes of the proposed
285~s and 142~s modes, the star would have to be viewed within
approximately 1$\degr$ of its equator. 
Because of this need for fine tuning and the inability to explain
the UV amplitudes, this model is inferior to the \el=4 hypothesis
we have presented.
 
\citet{WET} and \citet{Ke00} maintain that since the 142~s mode shows
the atypical behavior in the UV, it must be due to some nonlinear effect
and both the 71~s and 72~s modes are real modes.  As they point out,
a real mode with such short periods forces asteroseismological models
to use masses near the Chandrasekhar limit 
in order to create an \el=1, k=1 mode 
with such a low period \citep{B01}. 
As the mass of this star is near 0.6~M$_\odot$ \citep{WET}, 
they conclude these modes must be \el=2. They agree
with \citet{TC03} that the star must have an unfavorable
inclination to create the low amplitudes and possibly the 
character of the 142~s mode in the UV. The largest problem 
with this scenario comes from the nature of
the 142~s mode.  Combination and harmonic
modes are believed to be a result of nonlinear effects in the atmosphere
of the star.  However, the theories concerning the nonlinear
interaction between the outer atmosphere and the pulsation modes
predict the existence of harmonic modes but not 
sub-harmonics \citep{B92, Wu01}. The 142~s mode, 
as a subharmonic, would have to
be created by some new type of mechanism that operates
over every other cycle of the real, parent mode.
Unlike this scenario, our \el=4 hypothesis does 
not require the addition of any new physics. The
nonlinear effects discussed by \citet{B92} and \citet{Wu01}
act on the 142~s mode to create a harmonic near 71~s.

\section{Conclusions}

The DAV star \gone\ has a perplexing pulsation spectrum that has challenged
attempts to understand the star. We have analyzed line profile variations
that suggest the 142~s pulsation mode in \gone\ has a spherical degree of
four.  If this mode is the star's dominant pulsation mode, then we can
explain the star's overall low pulsation amplitude by geometric cancellation
without the need to invoke improbable inclinations.  Furthermore, because
the pulse shape harmonic of an \el=4 mode has lower \el\ characteristics, it
cancels less effectively, explaining the unusually large amplitude of the
mode's harmonic.  Finally, an \el=4 character of the 142~s mode and the
consequent \el=0,2 character of its harmonic are the very values most
consistent with the UV amplitudes measured by the Hubble Space Telescope,
although problems with pulsation phases remain.  Together, these results
yield a satisfying and unified picture of the star's pulsations, and the
only one consistent with all the data.

If our assignment of \el=4 to the 142~s mode is correct,
then it is a surprise since we have never before seen even \el=3
and only seldom \el=2.  Interestingly, the assignment of \el=1 to most of 
the modes in hot DAVs by \citet{C94} does not apply to the 142~s region
of \gone, which was one of two stars thought to have \el$>$1. 
Based on period spacings alone, \citet{C94} recognized that the 142~s
or its nearby 148~s mode must be higher \el\ but had suspected
\el=2, not 4. If we are forced to consider higher \el\ 
during mode identification, then it will make that process 
difficult because the high \el\ modes are closely space in 
period.  Fortunately, it appears that high \el\ modes advertise that 
fact by having large harmonics which may assist in their identification.

Finally, we have developed a new analysis technique for time series
spectroscopy that appears to work for relatively poor
signal-to-noise.  The other DAVs studied by time
series spectroscopy \citep{VK00,K02,K02b,K03} might
benefit from re-analysis by this technqiue, 
which we intend to do.  Also, we expect 
that by using this technique we will
be able to extend mode identification with
time-series spectroscopy to fainter stars, increasing
the number of stars for which secure mode identification is
possible.

\acknowledgments
Data presented herein were obtained at the W.M. Keck Observatory, 
which is operated as a scientific partnership among the California
Institute of Technology, the University of California and the National 
Aeronautics and Space Administration. The Observatory was made possible 
by the generous financial support of the W.M. Keck Foundation.
We would like to recognize the contribution
of the National Science Foundation through grant AST 000-94289 and
the support from the Alfred P. Sloan Foundation. S. Thompson
would like to thank the North Carolina Space Grant Consortium
for their financial assistance.

\end{document}